\documentclass[twoside,11pt]{article}

\usepackage{blindtext}
\usepackage{amsmath, amssymb, amsthm}
\usepackage{graphicx}
\usepackage{subcaption}
\usepackage{float}
\usepackage[a4paper, margin=1in]{geometry}
\usepackage{tikz}
\usetikzlibrary{arrows.meta,positioning}
\usepackage[colorlinks=true, urlcolor=blue, citecolor=blue]{hyperref}
\usepackage[backend=biber]{biblatex}
\usepackage{algorithm}
\usepackage{algpseudocode}
\def\Y{Y_\mathrm{obs}}

\newtheorem{theorem}{Theorem}[section]
\newtheorem{remark}{Remark}[section]

\title{Matrix Completion: Theory, Algorithms, and Empirical Evaluation}

\author{
\begin{tabular}{cc}
\textbf{Connor Panish} & \textbf{Leo Villani} \\
 cp732@cornell.edu & lv259@cornell.edu \\
Department of Statistics & Department of Statistics \\
Cornell University & Cornell University \\
Ithaca, NY & Ithaca, NY
\end{tabular}
}

\date{November 2025}

\begin{document}

\maketitle

\begin{abstract}
We present a concise survey of matrix completion methods and associated implementations of several fundamental algorithms. Our study covers both passive and adaptive strategies. We further illustrate the behavior of a simple adaptive sampling scheme through controlled synthetic experiments.
\end{abstract}

\section{Introduction}

We tackle the ubiquitous problem of \emph{matrix completion}: recovering a structured matrix from only a small subset of its observed entries. This problem appears in many modern applications, including recommender systems (user–item ratings, Netflix), collaborative filtering, and compressed sensing, where collecting all entries is either costly or impossible. To make recovery feasible, we assume the underlying matrix is low rank—a modeling choice that has natural interpretations in applications, such as latent user–item factors or low-dimensional signal structure.

\begin{figure}[h]
  \centering
  \resizebox{0.9\linewidth}{!}{
\begin{tikzpicture}[
    >=Latex,
    node distance=10mm and 12mm,
    block/.style={rectangle,rounded corners,draw,align=center,fill=gray!10,
                  minimum height=9mm,inner sep=2mm,text width=7.2cm},
    smallblock/.style={rectangle,rounded corners,draw,align=center,fill=gray!10,
                  minimum height=9mm,inner sep=2mm,text width=5.2cm}
]
\node[block] (rank) {%
$\displaystyle \min_{X\in\mathbb{R}^{n\times m}} \ \mathrm{rank}(X)
\quad \text{s.t.}\quad \mathcal{P}_{\Omega}(X)=\mathcal{P}_{\Omega}(Y)$\\[1mm]
\textbf{Projection constraint on observed entries}
};

\node[smallblock, below left=of rank] (altmin) {%
\textbf{Alternating estimation (factorization)}\\
$X=UV^\top,\ \min_{U,V}\ \|\mathcal{P}_{\Omega}(UV^\top - Y)\|_F^2$\\
(Alt-Min / ALS; nonconvex, scalable)
};
\node[smallblock, below right=of rank] (nuc) {%
\textbf{Nuclear-norm relaxation}\\
$\displaystyle \min_{X}\ \|X\|_* \quad \text{s.t.}\quad \mathcal{P}_{\Omega}(X)=\mathcal{P}_{\Omega}(Y)$\\
(convex surrogate for rank)
};

\draw[->,thick] (rank) -- (altmin);
\draw[->,thick] (rank) -- (nuc);

\node[smallblock, below left=of nuc] (sdp) {%
\textbf{SDP approach}\\
reformulate as a semidefinite program;\\
solve via interior point method: SDPT3
};
\node[smallblock, below right=of nuc] (svt) {%
\textbf{Iterative Singular Value Thresholding (SVT)}\\
ISTA with soft-thresholding on singular values
};

\draw[->,thick] (nuc) -- (sdp);
\draw[->,thick] (nuc) -- (svt);

\end{tikzpicture}}
  \caption{Roadmap of passive matrix completion methods.}
  \label{Passive_Methods}
\end{figure}
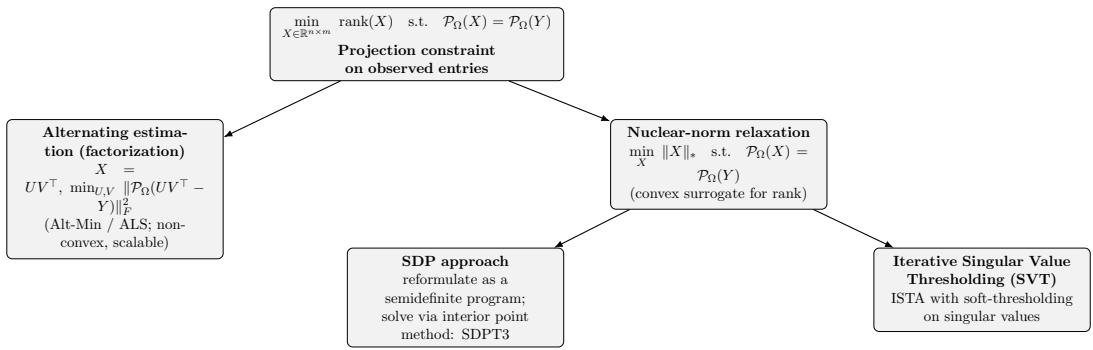

We study matrix completion under two paradigms, passive and adaptive. In the passive setting, one is given a single incomplete matrix at the beginning and asked to recover the matrix. The two key approaches are nuclear norm minimization and alternating minimization, see Figure \ref{Passive_Methods}. This setting mirrors the standard machine-learning paradigm in which a fixed dataset is provided and the goal is to infer parameters or structure that generalize to new data.

In contrast, adaptive learning methods consider a dynamic setting in which the learner can iteratively choose which entries to observe next. Here, the flow of data is influenced by the algorithm itself: based on the current partial information, the method selects informative entries to query, often according to a scoring or uncertainty criterion. This paradigm better reflects real-world systems where measurements are expensive and must be acquired in some strategic way.

\subsection*{Problem Setup}
We begin with a partially observed data matrix $Y \in \mathbb{R}^{n\times m}$ together with an index set
\[
\Omega \subset [n]\times[m]
\]
of observed entries. Let $\mathcal{P}_{\Omega}:\mathbb{R}^{n\times m}\to\mathbb{R}^{n\times m}$ denote the projection operator
\[
\big(\mathcal{P}_\Omega(X)\big)_{ij}
=
\begin{cases}
X_{ij}, & (i,j)\in\Omega,\\
0, & (i,j)\notin\Omega.
\end{cases}
\]
where $\Y := \mathcal{P}_{\Omega}(Y)$ denotes the observed matrix, with 0's in the missing entries. In order to make matrix completion possible, we assume the true matrix $Y$ is low-rank with rank $r \ll \min(n,m)$. Furthermore, note that the low rank assumption is not enough. Let $e_1e_n^T$ be the standard basis vectors, this matrix is low rank, but is infeasible to recover without observing all the entries. Therefore, we also often assume the standard incoherence condition with parameter $\mu_0$, that is for the SVD of $Y = U \Sigma V^T$ we have
\begin{equation}
    \label{Incoherence_Assumption}
    \begin{aligned}
        \max_{1 \leq i \leq n} ||U^Te_i||_2 &\leq \sqrt{\frac{\mu_0r}{n}} \\
        \max_{1 \leq j \leq m} ||V^Te_j||_2 &\leq \sqrt{\frac{\mu_0r}{m}}
    \end{aligned}
\end{equation}
where $1 \leq \mu_0 \leq \frac{\min(n,m)}{r}$ \cite{Chen15}. This condition ensures that the values of $Y$ are equally spread out over the rows and columns, ruling out the pathological example given above. In the passive setting, incoherence is often assumed while adaptive methods allow for this condition to be relaxed. In our initial experiments we work in a synthetic setting where $Y$ is generated as a low rank matrix 
\[
Y = UV^T, \quad U \in \mathbb{R}^{n\times r}, V \in \mathbb{R}^{m\times r}
\]
and the observation pattern is missing completely at random (MCAR); each entry is included independently with probability $p_{\mathrm{obs}}$. Later sections relax these assumptions by introducing entry-dependent costs and noisy observations. This paper is primarily expository: our goal is to consolidate classical results on passive and adaptive matrix completion, and to compare a few canonical algorithms in a typical experimental framework.\\
The fundamental goal is to recover a good or ideally a perfect approximation of $Y$, denoted $\hat{Y}$ from $\Y$.

\section{Passive Methods}
The ideal mathematical formulation would be 
\begin{equation}
\label{eq:rank-min}
\min_{X\in\mathbb{R}^{n\times m}} \ \mathrm{rank}(X)
\quad \text{s.t.}\quad
\mathcal{P}_\Omega(X) = \mathcal{P}_\Omega(Y),
\end{equation}
but this problem is NP-hard in general. The standard convex surrogate is to replace the rank with the nuclear norm $\|X\|_* = \sum_i \sigma_i(X)$:
\begin{equation}
\label{eq:nuc-constraint}
\min_{X} \ \|X\|_*
\quad\text{s.t.}\quad
\mathcal{P}_\Omega(X) = \mathcal{P}_\Omega(Y),
\end{equation}

\subsection{Semidefinite Program Formulation}

The nuclear norm minimization problem \ref{eq:nuc-constraint} can be reformulated as a semidefinite program in terms of the trace
$$
\text{Minimize} \quad \frac{1}{2} \left( \text{Tr}(W_1) + \text{Tr}(W_2) \right)
$$
$$
\text{Subject to:} \quad
\begin{bmatrix}
W_1 & X \\
X^T & W_2
\end{bmatrix} \succeq 0
$$
$$
X_{ij} = Y_{ij} \quad \text{for all } (i, j) \in \Omega
$$
for $W_1 \in \mathbb{R}^{n\times n}$ and $W_2 \in \mathbb{R}^{m \times m}$ which is given in \cite{Candes08} and proved in Chapter 5 of Fazel's PhD Thesis \cite{Fazel02}. Intuitively, note that for a symmetric positive definite matrix the nuclear norm is the trace. This approach is extremely computationally expensive, due to the high number of constraints and number of variables, and hence takes minutes for even $100\times 100$ matrices. We will compare this direct SDP approach against the much more scalable iterative SVT method.\\

\subsection{Singular Value Thresholding}

To begin, we relax the nuclear norm minimization problem to
\begin{equation}
\label{SVT_Norm}
\min \tau \|X\|_* + \frac{1}{2}\|X\|^2_F
\quad\text{s.t.}\quad
\mathcal{P}_\Omega(X) = \mathcal{P}_\Omega(Y),
\end{equation}
where $\tau$ is a tuning parameter. This minimization problem can be solved using singular value thresholding on the SVD of $X$. Define
\[D_{\tau}(X) = UD_{\tau}(\Sigma)V^T \hspace{2em} D_{\tau}(\Sigma) = \text{diag}(\{(\sigma_i-\tau)_+\})\]
which is the singular value thresholding of the diagonal matrix. Now for the algorithm to solve~\eqref{SVT_Norm}, let $\tau$ be fixed and $\{\delta_k\}$ be a set of step sizes. Initializing with $U_0 = \Y$,
\begin{align*}
    \text{for }k &= 1,2,3,... \\
    X^k &= D_{\tau}(U_{k-1}) \\
    U^k &= U^{k-1} + \delta_k P_{\Omega}(Y - X^k)
\end{align*}
This algorithm is outlined in \cite{Cai08}. In practice, we set a maximum number of iterations and set the stopping condition
\begin{equation*}
    \frac{\|P_{\Omega}(Y - X^k)\|_F}{\|P_{\Omega}(Y)\|} = \frac{\|P_{\Omega}(Y - X^k)\|_F}{\|\Y\|} < \epsilon 
\end{equation*}
where we set $\epsilon = 10^{-4}$. Furthermore, $\tau$ should be sufficiently large in order for nuclear norm minimization~\eqref{eq:nuc-constraint} to be approximated by~\eqref{SVT_Norm}. We set $\tau = \frac{5(n+m)}{2}$ which is justified in \cite{Cai08} through a heuristic argument. In addition, the method will converge when the step sizes $0 < \delta_k < 2$, we simply set $\delta_k = 1$, but occasionally in order to speed up convergence you choose a larger step size $\delta_k  = 1.2\frac{nm}{|\Omega|}$. Finally, for implementation purposes there are additional details related to the optimal way to perform singular value thresholding which are mentioned in \cite{Cai08}.

\subsection{Alternating Minimization}
A common nonconvex approach to matrix completion is to factor the unknown
matrix as
\[
X = UV^\top,\qquad U \in \mathbb{R}^{n\times r},\ V \in \mathbb{R}^{m\times r}.
\]
This factorization can be viewed as a latent-factor model: each row $i$ has a latent vector
$u_i \in \mathbb{R}^r$ and each column $j$ a latent vector $v_j \in \mathbb{R}^r$, so that
\[
Y_{ij} \approx X_{ij} = u_i^\top v_j.
\]
Thus, the entry at $(i,j)$ depends only on the latent features of its row and column. More complex
models replace the inner product by a nonlinear interaction $f(u_i, v_j)$, but still assume that all
dependence is mediated by row and column factors.

Given this parameterization, we directly minimize the squared error on the observed entries:
\begin{equation}
\label{eq:altmin-obj}
\min_{U,V}\ 
\|\mathcal{P}_\Omega(UV^\top - Y_{\mathrm{obs}})\|_F^2.
\end{equation}

For fixed $V$, the objective in~\eqref{eq:altmin-obj} is quadratic in $U$ and
decouples row-wise into least-squares problems. Similarly, for fixed $U$
it is quadratic in $V$ and decouples column-wise. Alternating minimization
(AltMin) exploits this structure by iteratively solving these least-squares
subproblems. A derivation showing this
decoupling is provided in Appendix~\ref{app:altmin-decoupling}.

This method is very computationally cheap: decoupling the problem reduces in some sense the relative cost of each step. The drawbacks are its sensitivity to initialization and the need to specify (or estimate) the underlying rank. Poor initialization of $U^{(0)}, V^{(0)}$ can lead to falling into bad local minima so in practice we compute a rank-$r$ truncated SVD of the masked matrix
$\Y$ and initialize
\[
M \approx U_0 \Sigma_0 V_0^\top,\qquad
U^{(0)} := U_0,\quad
V^{(0)} := V_0 \Sigma_0.
\]

\paragraph{Theoretical guarantees.}
Although the optimization problem~\eqref{eq:altmin-obj} is nonconvex, it also admits rigorous recovery guarantees under assumptions similar to those used for nuclear norm minimization. Several works (e.g.\ Jain et al. \cite{jain2012}) show that, under uniform sampling and incoherence, alternating minimization with a good spectral initialization recovers the true matrix with high probability using a comparable number of samples.

\begin{theorem}[Alternating minimization, informal]
\label{thm:altmin}
Assume $Y$ is rank-$r$ and $\mu_0$-incoherent, and that the observed entries $\Omega$ are sampled uniformly at random with
\[
|\Omega| \;\gtrsim\; \mu_0\, r\, n \,\mathrm{polylog}(n).
\]
If the initial factors $(U^{(0)}, V^{(0)})$ are sufficiently close to the true singular vectors (for example, obtained from a truncated SVD of $\Y$), then the alternating minimization iterates converge geometrically to $Y$ with high probability.
\end{theorem}

\begin{figure}[H]
\centering
\begin{tabular}{ccc}
\includegraphics[width=0.30\linewidth]{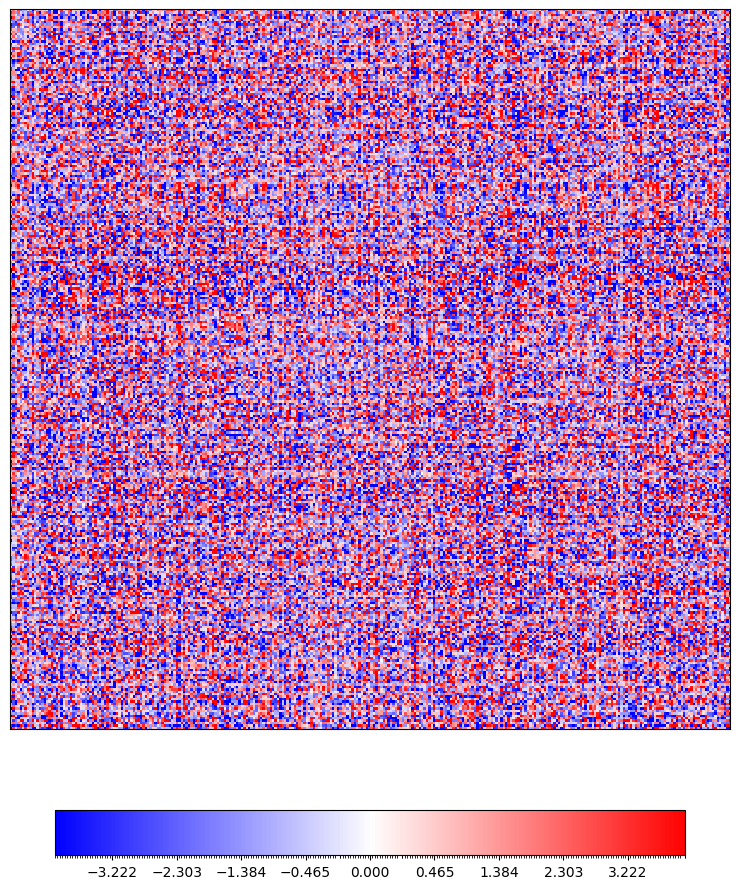} &
\includegraphics[width=0.30\linewidth]{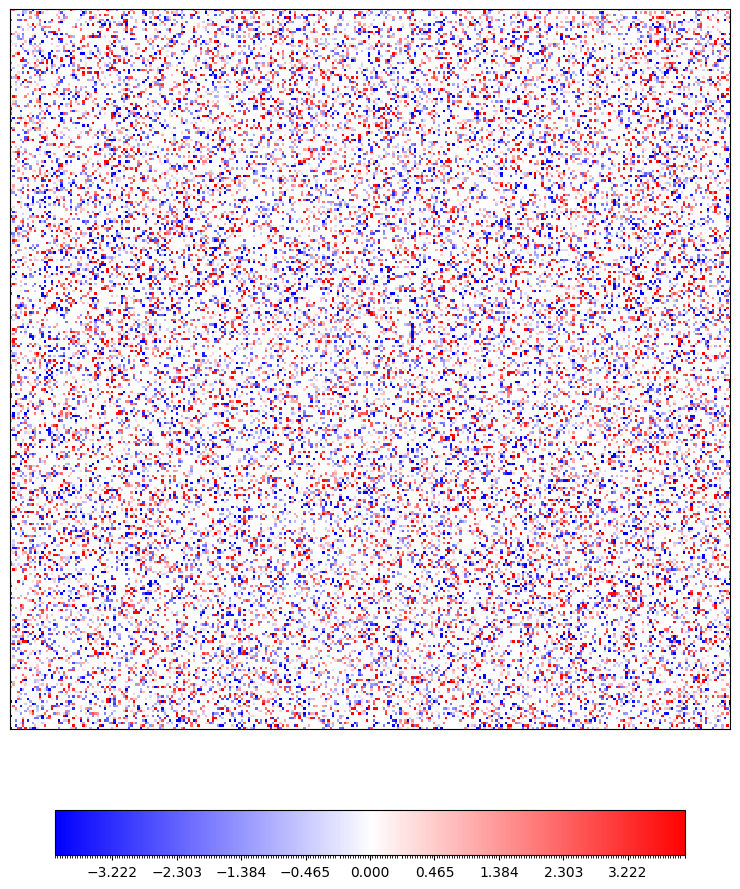} &
\includegraphics[width=0.30\linewidth]{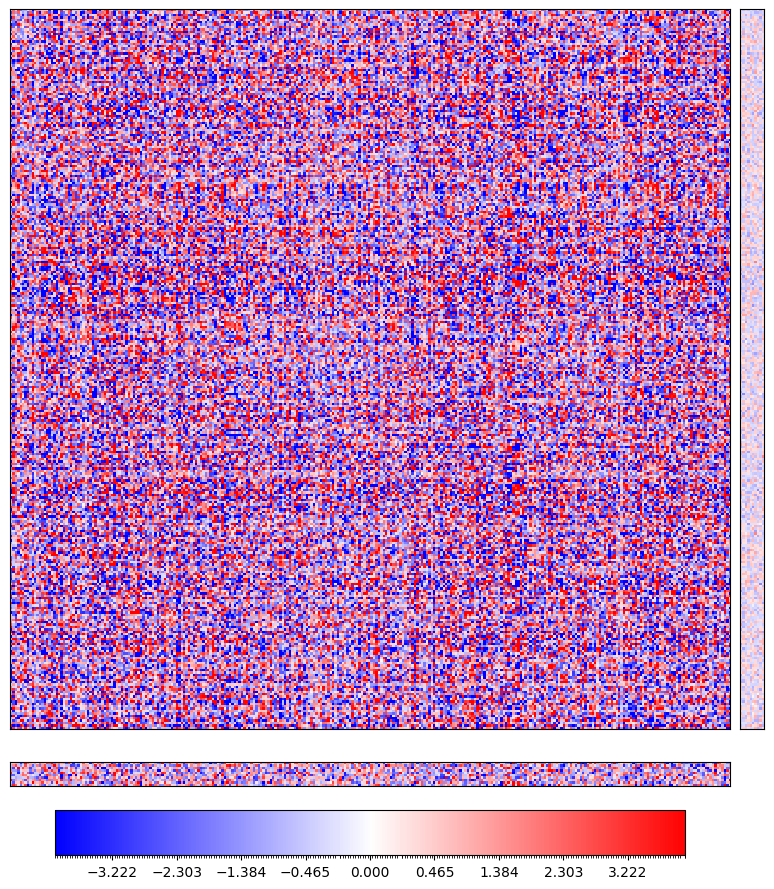} \\
(a) True low-rank matrix 
& (b) Observed entries
& (c) AltMin after 10 iterations
\end{tabular}
\caption{
Visualization of alternating minimization on a synthetic matrix completion instance 
with $n=m=300$, $p_{\textrm{obs}} = 0.3$ and rank $r=10$.  
Despite starting from incomplete data, the algorithm rapidly converges and 
recovers the global structure of the matrix.
}
\label{fig:altmin-visualization}
\end{figure}

Our initialization scheme based on the rank-$r$ truncated SVD of $\Y$ matches this theoretical template, and our experiments in Figure~\ref{fig:altmin-visualization} illustrate the resulting rapid convergence in practice.

\subsection{Experiments / Comparisons}

\begin{figure}[H]
\centering
\begin{subfigure}{0.4\linewidth}
    \centering
    \includegraphics[width=\linewidth]{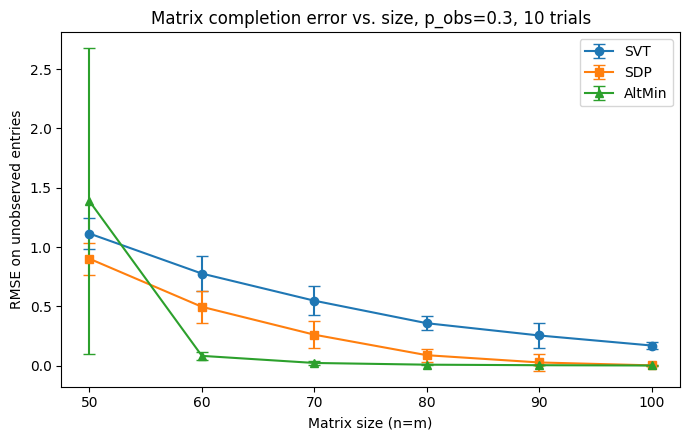}
    \caption{RMSE vs.\ matrix size ($n=m$)\\ \,}
    \label{fig:rmse-vs-n}
\end{subfigure}
\hfill
\begin{subfigure}{0.4\linewidth}
    \centering
    \includegraphics[width=\linewidth]{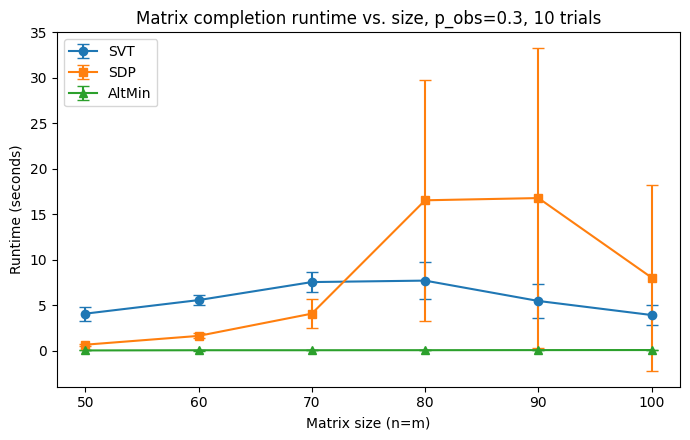}
    \caption{Runtime vs.\ matrix size. AltMin scales well, while SDP and SVT degrade.}
    \label{fig:time-vs-n}
\end{subfigure}

\caption{Comparison of RMSE and runtime across matrix sizes for SDP, SVT, and AltMin.
Each point represents the average over 10 independent trials with error bars indicating
one standard deviation. Note runtime decreases for SDP, since the number of observations $.3n^2$ is growing faster then the theoretical rate of $nr\log^2n$ in~\eqref{thm:exact-nuc}. Therefore, the matrix completion problem is getting easier with increased size.}
\label{fig:rmse-time-vs-n}
\end{figure}

For each matrix size, we report the mean and one standard deviation of both RMSE
and runtime over 10 independent trials. The results are summarized in
Figure~\ref{fig:rmse-time-vs-n}.

\begin{figure}[H]
\centering
\begin{subfigure}{0.38\linewidth}
    \centering
    \includegraphics[width=\linewidth]{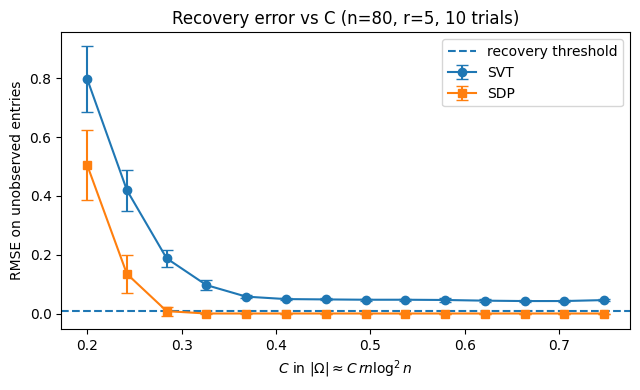}
    \caption{Number of Samples vs. RMSE to recover $C$}
    \label{fig:errorvsC}
\end{subfigure}
\hfill
\begin{subfigure}{0.38\linewidth}
    \centering
    \includegraphics[width=\linewidth]{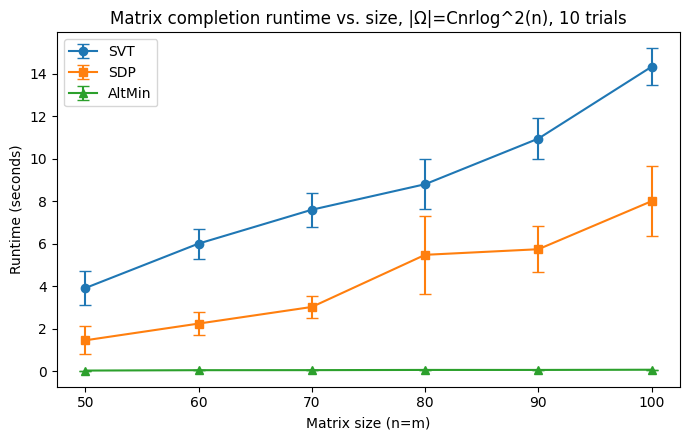}
    \caption{Runtime plot with consistent hardness}
    \label{fig:timevslog2n}
\end{subfigure}
\end{figure}

To better understand how our methods behave relative to the theoretical sampling rate, we also do a sweep over the constant multiplier $C$ in the predicted sample complexity $Crn\log^2n$. Figure~\ref{fig:errorvsC} shows the empirical RMSE for SVT and SDP as $C$ increases. Both methods exhibit a clear phase transition: once the number of samples exceeds a modest constant multiple of $nr\log^2n$, recovery becomes essentially exact.\\
To control for the progressive oversampling in the previous runtime plot we include in Figure~\ref{fig:timevslog2n} a ``consistent hardness'' experiment in which each matrix is sampled at the theoretical rate. This removes the confounder of easier instances at larger $n$ yielding a runtime plot that increases monotonically.\\

\begin{figure}[H]
    \centering
    \includegraphics[width=0.7\linewidth]{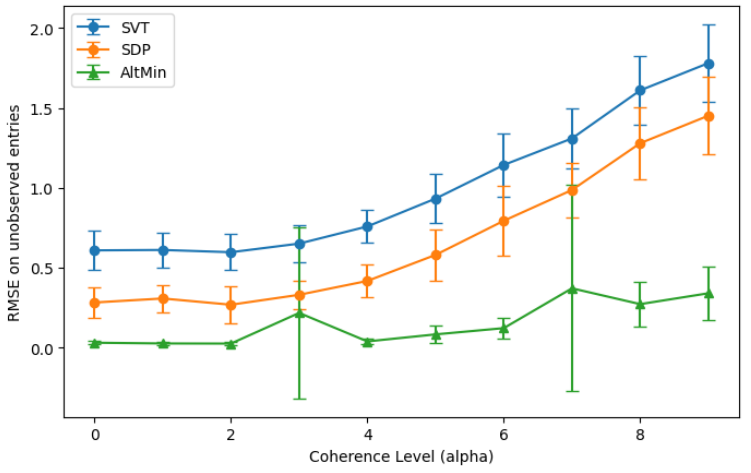}
    \caption{Comparison of RMSE and coherence across varying coherence for SDP, SVT, and AltMin. Each point represents the average over 10 independent trials with error bars indicating one standard deviation.}
    \label{fig:passive-coherence}
\end{figure}

\noindent{}RMSE was computed via:
\[
\mathrm{RMSE}(M, \hat{M}) = \sqrt{\frac{1}{|\bar{\Omega}|} \sum_{(i,j) \in \bar{\Omega}} (M_{ij} - \hat{M}_{i,j})^2}
\]
or simply the Frobenius norm over the unseen entries.\\
A hallmark of passive methods is breaking down for high coherence. There are two main ways to construct coherent matrices, the first is power law matrices constructed as $DUV^TD$ where $D$ is diagonal and $D_{ii} = i^{-\alpha}$. The second is adding on block matrices, for~\eqref{Incoherence_Assumption} we only add on row coherence by letting
\[A = U + \alpha \begin{bmatrix}
    I_r & 0 \\ 0 & 0
\end{bmatrix}\]
and performing matrix completion on $AV^T$. Similarly, one could add column coherence if desired.
The results are summarized in
Figure~\ref{fig:passive-coherence}.

We also wanted to evaluate on the real MovieLens 100K dataset. Here is the heatmap for $200\times 200$ subset of the matrix and the full matrix where rating $0$ indicates missing:
\begin{figure}[h!]
    \centering
    
    \begin{subfigure}{0.47\textwidth}
        \centering
        \includegraphics[width=\linewidth]{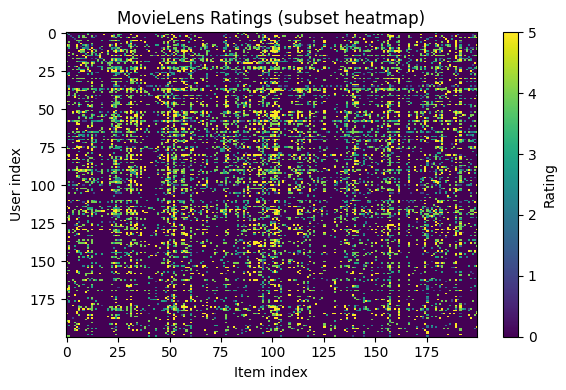}
        \caption{$200\times 200$ subset}
    \end{subfigure}
    \hfill
    \begin{subfigure}{0.47\textwidth}
        \centering
        \includegraphics[width=\linewidth]{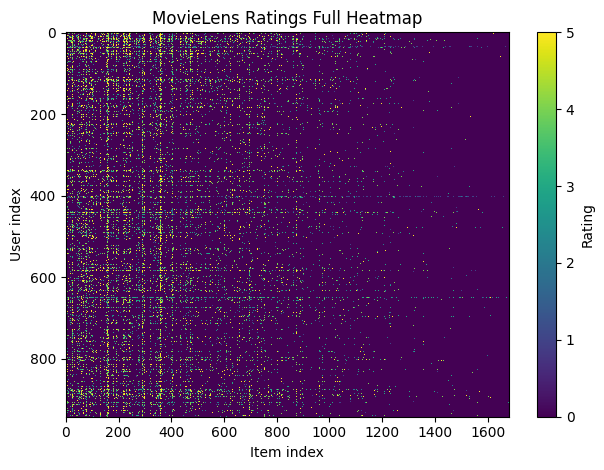}
        \caption{Full MovieLens matrix}
    \end{subfigure}
    
    \caption{Heatmaps of MovieLens ratings: a small subset vs. the full user–item matrix.}
    \label{fig:movielens-heatmaps}
\end{figure}

\begin{figure}[H]
    \centering
    \includegraphics[width=0.7\linewidth]{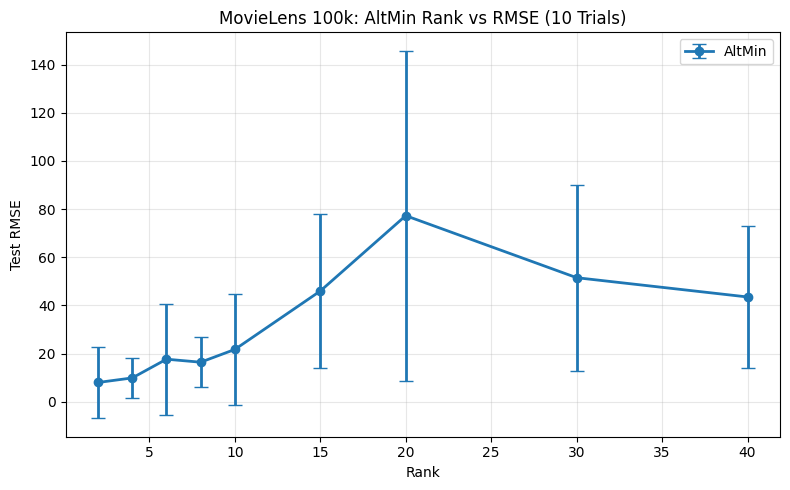}
    \caption{Cross-validated test RMSE on MovieLens 100K as a function of the rank $r$ in the AltMin factorization.}
    \label{fig:movielens-crossvalid}
\end{figure}

One large problem is SDP and SVT become computationally prohibitive on the full MovieLens 100K dataset in our implementation. SDP requires solving a large semidefinite program, and SVT performs repeated SVDs on a dense $943\times 1682$ matrix. In contrast, AltMin only solves many small least-squares problems and scales easily to MovieLens. For this reason, our real-data experiments on MovieLens focus on AltMin. We tune the rank $r$ by cross-validation and observe that the test RMSE is minimized for a relatively small value of $r$, as shown in Figure~\ref{fig:movielens-crossvalid}.

This agrees with prior work on the Netflix Prize dataset, where only 5–20 latent factors are typically needed to model the underlying preference structure.

\subsection{Some Theory}
For completeness, we briefly recall the classical guarantees for nuclear-norm based matrix completion. Let $Y \in \mathbb{R}^{n_1 \times n_2}$ be a rank-$r$ matrix, and write $n = \max(n_1,n_2)$. We observe a uniformly random subset of entries $\Omega \subset [n_1]\times[n_2]$, and consider the nuclear norm program~\eqref{eq:nuc-constraint}.

\begin{theorem}[Exact recovery via nuclear norm, informal]
\label{thm:exact-nuc}
Assume that $Y$ has rank $r$ and satisfies a $\mu_0$-incoherence condition on its row and column spaces. If the observed set $\Omega$ is drawn uniformly at random and
\[
|\Omega| \;\gtrsim\; \mu_0\, r\, n \log^2 n,
\]
then, with high probability, the solution of the nuclear norm program~\eqref{eq:nuc-constraint} is unique and equals $Y$.
\end{theorem}

This informal statement summarizes the results of Cand\`es and Recht~\cite{Candes08} and Recht~\cite{Recht11}. Up to logarithmic factors, the required number of samples is essentially optimal.

\begin{remark}[Near-optimality]
\label{rem:lower-bound}
Cand\`es and Tao~\cite{CandesTao10} show that, under uniform sampling, if
\(
|\Omega|
\)
is smaller than a constant multiple of $\mu_0\, r\, n \log n$, then there exist rank-$r$ matrices that cannot be recovered by any nuclear-norm based method. Modulo log factors, the sample complexity
\[
|\Omega| \asymp \mu_0\, r\, n
\]
is therefore information-theoretically sharp.
\end{remark}

We also note that nuclear norm minimization is stable to noise. Suppose the observed entries are corrupted by additive noise:
\[
M_{ij} = Y_{ij} + Z_{ij}, \qquad (i,j)\in\Omega,
\]
where $Y_{ij}$ is the true rank-$r$ matrix and $Z_{ij}$ are noise terms. Letting $Z$ be the noise matrix, assuming that $\|\mathcal{P}_\Omega(Z)\|_F \le \delta$. Given this, we have that the solution $X$ to the nuclear norm program
\[\min_X||X||_* \text{ s.t. } ||\mathcal{P}_{\Omega}(X-M)||_F \leq \delta\]
has the property that
\[\|Y - X\|_F \le 4\delta \sqrt{\frac{C_p \min(n,m)}{p}} + 2\delta.
\]
with high probability where $p$ denotes the sampling probability and $C_p$ is an absolute constant (see \cite{candes2009noise} for details).

In our synthetic experiments we take $n_1 = n_2 = n$ and observe each entry independently with probability $p_{\mathrm{obs}} = 0.3$, so that
\[
|\Omega| \approx 0.3\, n^2.
\]
Comparing this to the theoretical requirement $|\Omega| \gtrsim \mu_0\, r\, n \log n$,
we see that
\[
\frac{|\Omega|}{\mu_0\, r\, n \log n}
\;\approx\;
\frac{0.3}{\mu_0}\,\frac{n}{r \log n},
\]
which grows with $n$ for fixed rank $r$ and moderate $\mu_0$. Thus, as we increase
the matrix size while keeping $r$ and $p_{\mathrm{obs}}$ fixed, we become
increasingly oversampled relative to the theoretical threshold, and recovery should become easier. This agrees with our empirical results in Figure~\ref{fig:rmse-vs-n}: for larger $n$, both SDP and SVT achieve lower RMSE and AltMin converges more reliably.

\section{Adaptive Methods}

The key limitation of passive methods is their reliance on an incoherence assumption~\eqref{Incoherence_Assumption}, the entries are evenly spread out. In the adaptive setting, we assume that we can ask an oracle for entries based upon the entries we initially reveal.

\subsection{Tensor Completion}
For this method developed by Krishnamurthy and Singh \cite{Krishnamurthy13}, we relax the condition of row coherence, and maintain column coherence. We begin by randomly observing a proportion of the total rows $p_{\mathrm{row}}$, which defines our observed entries $\Omega$. Note unlike the previous setting, we are observing multiple complete rows.

The algorithm is as follows, observe a proportion $p_{\mathrm{row}}$ of the rows of the matrix $Y$. We then use these observations, to iteratively build a column space $\Tilde{\mathcal{U}}$. If we determine the column is in the columns space $\Tilde{\mathcal{U}}$ then we reconstruct it using least squares. Otherwise, we fully observe the column and add it to our column space.
In practice we set a tolerance $\varepsilon$ which controls whether a partially observed column is considered to lie within the current estimate of the subspace. This allows recovery even when row incoherence fails, as long as the columns remain incoherent we only to fully reveal a modest number of columns. In many cases, the number of fully revealed columns is exactly the rank of the underlying matrix. \\

\begin{figure}[H]
    \centering
    \includegraphics[width=0.7\linewidth]{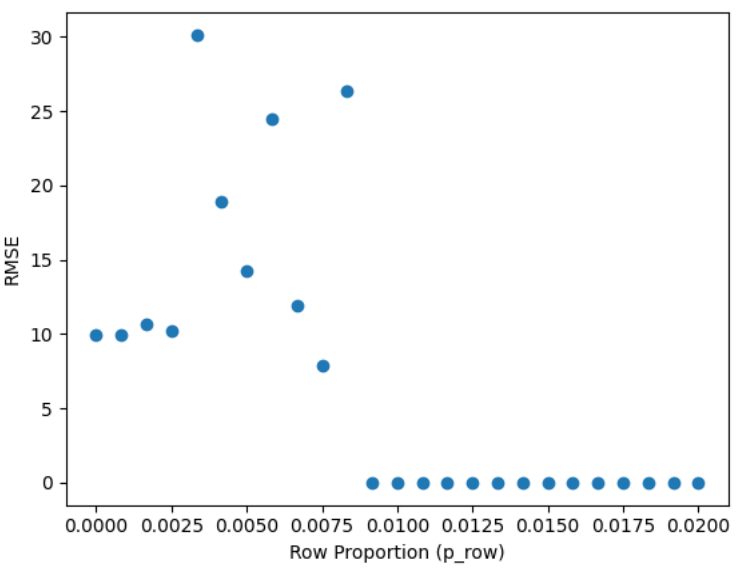}
    \caption{Row proportion vs RMSE on a high coherence matrix}
    \label{fig:tensor-comp}
\end{figure}

Figure~\ref{fig:tensor-comp} summarizes the relationship between proportion of entries revealed vs. RMSE for a matrix with high row coherence. This is on a highly row coherent $500 \times 500$ Gaussian matrix.

\subsection{Leverage-Based Adaptive Sampling}
A more principled adaptive strategy is based on the notion of \emph{leverage scores}, introduced for matrix completion by Chen et al.~\cite{Chen14}. The key insight is that recovery is determined by how concentrated the singular vectors are. When certain rows and columns are highly coherent uniform sampling is unlikely to observe them and passive methods fail. Leverage based methods address this via two steps:
\begin{enumerate}
    \item First sample a small number of entries uniformly at random and compute a rough estimate of the row and column subspaces via a truncated SVD. The goal is to quickly identify the most informative directions of the matrix.
    \item  In the second phase, sampling becomes \emph{non-uniform}: each unobserved entry is sampled with probability proportional to the sum of its row and column leverage scores.
\end{enumerate}
With this targeted sampling, one can succeed in many regimes with essentially $O(rn)$ samples, as opposed to the typical $O(\mu_0 rn \log n)$  rate needed under uniform sampling. Intuitively leverage-based sampling identifies the ``spiky'' parts of the matrix and then focuses measurements there making it simple and effective.

\section{Other Adaptive Experiments}
In addition to previous adaptive based sampling methods, we consider a simple toy model in which adaptivity is driven by a notion of \emph{uncertainty} together with
entry-dependent \emph{costs}. The goal is not to recover the matrix as accurately as possible or to be optimal in any sense, but rather to illustrate how different cost structures interact with an uncertainty-based acquisition heuristic.

\subsection{Toy Uncertainty--Cost Model}
We generate a noisy low-rank matrix
\[
M = U V^\top \in \mathbb{R}^{n\times m}, \qquad
Y = M + \sigma Z,
\]
where $U \in \mathbb{R}^{n\times r}$, $V \in \mathbb{R}^{m\times r}$ have i.i.d.\ Gaussian
entries and $Z\sim \mathcal{N}(0,1)$. We then compute 
\[
f_{\mathrm{row}}(i) = 1 -\frac{\#\{j : \mathrm{\Omega}_{ij}=1\}}{m}, \quad f_{\mathrm{col}}(j) = 1 -\frac{\#\{i : \mathrm{\Omega}_{ij}=1\}}{n}
\]
and for each entry an uncertainty score:
\[
S_{ij} = \frac{1}{2}\Big(
f_{\mathrm{row}}(i) + f_{\mathrm{col}}(j)
\Big),
\]
which is large when both row $i$ and column $j$ are undersampled. We set $S_{ij} = 0$ for $(i,j) \in \Omega$. In addition each entry has a cost, $C_{ij} > 0$ which models the expense of querying the entry. We then compute utility scores:
\[
U_{ij} = \frac{S_{ij}}{C_{ij}}
\]
so that high uncertainty, low cost scores are preferred. At each adaptive step we reveal the entry with maximum utility,
\[
(i_t,j_t) = \arg\max_{(i,j) \notin \Omega} U_{ij},
\]
update the mask, and recompute the row/column uncertainty and utility
scores. We also track the cumulative cost
\[
\mathrm{cost}(T) = \sum_{t=1}^T C_{i_t j_t}
\]
\subsection{Cost Models}
\begin{enumerate}
    \item[C1.] \textbf{Random i.i.d.\ costs.}
    Each entry has an independent random cost
    $C_{ij} \sim \mathrm{Unif}[0.5,1.5]$.  
    In this case there is essentially no structure in the cost landscape, so the
    acquisition rule is driven almost entirely by the uncertainty scores.

    \item[C2.] \textbf{Expensive users (row structure).}
    Most entries are inexpensive, but a fixed subset of rows is uniformly costly:
    \begin{itemize}
        \item baseline cost $C_{ij} \approx 0.5$ for most entries;
        \item $10\%$ of rows are designated ``expensive,'' with all entries in those
        rows set to a much larger cost (e.g.\ $C_{ij} = 10$).
    \end{itemize}
    Under this model we expect the algorithm to avoid these rows initially, only
    querying them when their uncertainty becomes so large that it outweighs their
    cost.

    \item[C3.] \textbf{Correlated structure (frontier model).}
    In this final scenario we imagine a world in which ``new'' users or
    ``new'' item categories live toward the bottom-right of the matrix, and both
    uncertainty and cost increase in that direction.  
    We model this with a cost that grows along a diagonal frontier:
    \[
    C_{ij} \approx 0.1 + 2\!\left(\frac{i}{n} + \frac{j}{m}\right),
    \]
    plus a small random perturbation.  
    We also bias the initial mask toward the top-left (``older'' data) by sampling
    $(i,j)$ with higher probability when $i$ and $j$ are small.  
    This creates a setting where the bottom-right region is simultaneously
    highly uncertain and expensive to query.
\end{enumerate}
We can optionally introduce dynamic pricing effects, though we do not include the corresponding plots in this report. Two simple variants are:
\begin{itemize}
    \item \emph{Surge pricing:} after querying $(i,j)$, all costs in row $i$ are
    multiplied by a factor $>1$, making future queries in that row more expensive.
    \item \emph{Bulk discount:} after querying $(i,j)$, all costs in row $i$ are
    multiplied by a factor $<1$, encouraging additional queries in that row.
\end{itemize}
These simple mechanisms let us visualize how the focus of the heuristic acquisition strategy shifts in response to both uncertainty and changing costs.
\subsection{Experiments}
\begin{figure}[H]
\centering
\begin{subfigure}{0.32\linewidth}
    \centering
    \includegraphics[width=\linewidth]{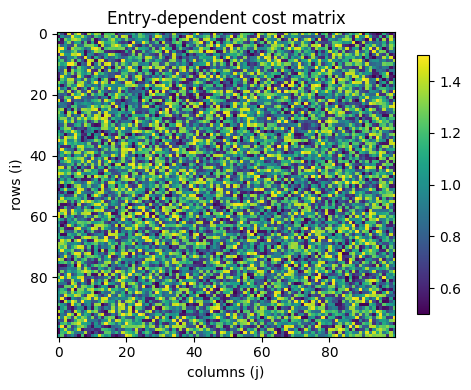}
    \caption{Cost matrix $C$.}
\end{subfigure}
\hfill
\begin{subfigure}{0.32\linewidth}
    \centering
    \includegraphics[width=\linewidth]{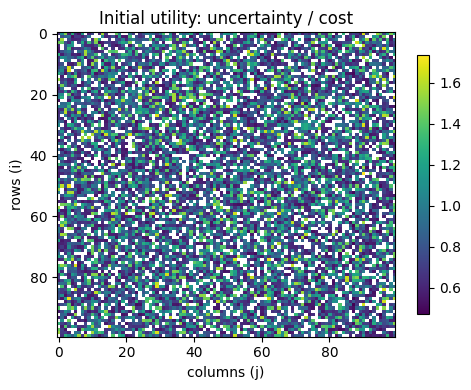}
    \caption{Initial utility $U = S_0 / C$.}
\end{subfigure}
\hfill
\begin{subfigure}{0.32\linewidth}
    \centering
    \includegraphics[width=\linewidth]{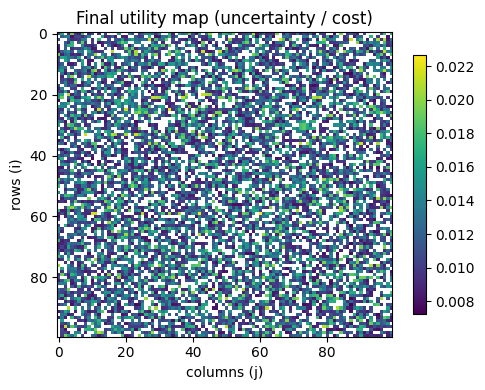}
    \caption{Final utility after $T$ reveals.}
\end{subfigure}
\caption{
Random-cost setting (Model C1). 
}
\label{fig:c1-maps}
\end{figure}

Experiments below will be run on a $n=m=100$ size matrix with true rank, $r = 5$, $p_{\mathrm{obs}} = 0.2$ and $\sigma = 0.2$.\\
For Model C1 we run the uncertainty strategy for $T=1000$ reveals.
Figure~\ref{fig:c1-maps} summarizes the geometry of the problem and the effect of the policy.

\begin{figure}[H]
\centering
\begin{subfigure}{0.32\linewidth}
    \centering
    \includegraphics[width=\linewidth]{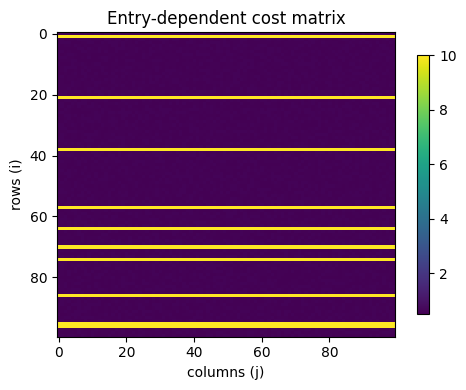}
    \caption{Cost matrix $C$.}
\end{subfigure}
\hfill
\begin{subfigure}{0.32\linewidth}
    \centering
    \includegraphics[width=\linewidth]{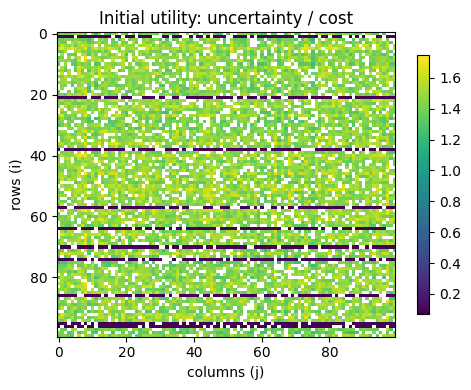}
    \caption{Initial utility $U = S_0 / C$.}
\end{subfigure}
\hfill
\begin{subfigure}{0.32\linewidth}
    \centering
    \includegraphics[width=\linewidth]{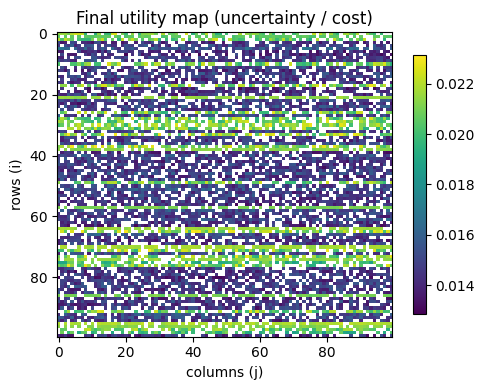}
    \caption{Final utility after $T$ reveals.}
\end{subfigure}
\caption{
Row Structured setting (Model C2). 
}
\label{fig:c2-maps}
\end{figure}

Similarly, Figure~\ref{fig:c2-maps} summarizes Model C2.

\begin{figure}[H]
\centering
\begin{subfigure}{0.24\linewidth}
    \centering
    \includegraphics[width=\linewidth]{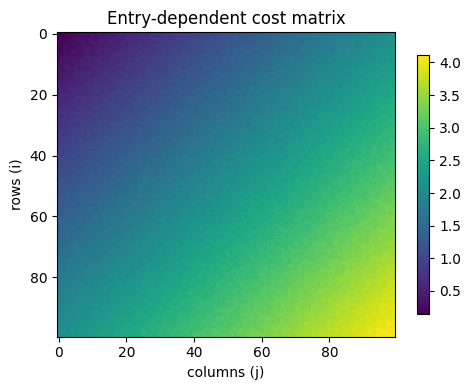}
    \caption{Cost matrix $C$.}
\end{subfigure}
\hfill
\begin{subfigure}{0.24\linewidth}
    \centering
    \includegraphics[width=\linewidth]{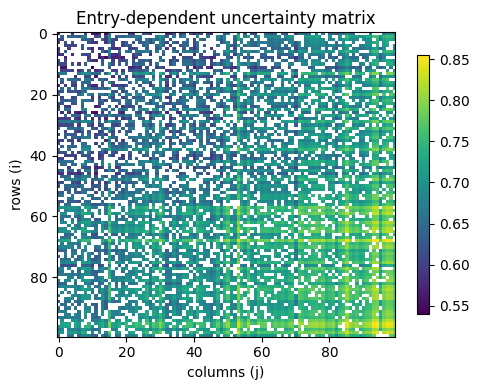}
    \caption{Initial Uncertainty}
\end{subfigure}
\hfill
\begin{subfigure}{0.24\linewidth}
    \centering
    \includegraphics[width=\linewidth]{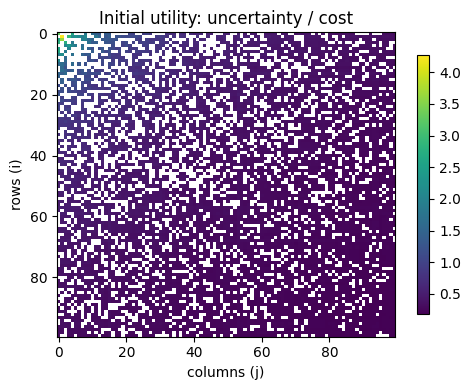}
    \caption{Initial utility}
\end{subfigure}
\hfill
\begin{subfigure}{0.24\linewidth}
    \centering
    \includegraphics[width=\linewidth]{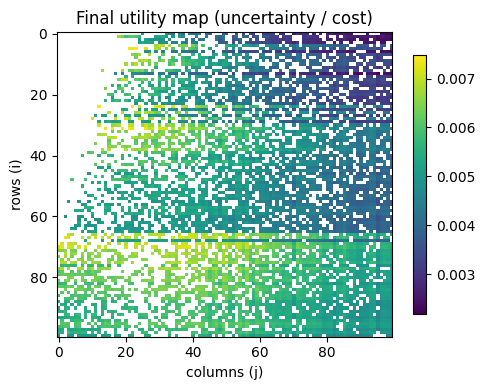}
    \caption{Final utility}
\end{subfigure}
\caption{
Correlated Structure (Model C3). 
}
\label{fig:c3-maps}
\end{figure}

Figure~\ref{fig:c3-maps} shows the correlated structure (Model C3).

\begin{figure}[H]
\centering
\begin{subfigure}{0.32\linewidth}
    \centering
    \includegraphics[width=\linewidth]{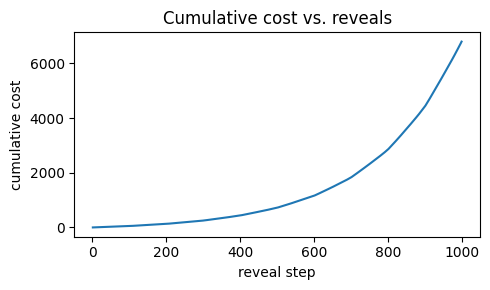}
    \caption{C1}
\end{subfigure}
\hfill
\begin{subfigure}{0.32\linewidth}
    \centering
    \includegraphics[width=\linewidth]{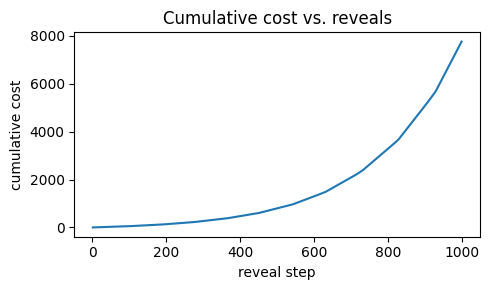}
    \caption{C2}
\end{subfigure}
\hfill
\begin{subfigure}{0.32\linewidth}
    \centering
    \includegraphics[width=\linewidth]{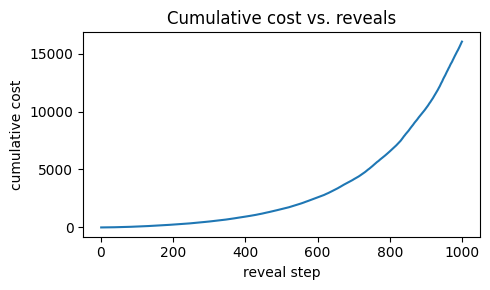}
    \caption{C3}
\end{subfigure}
\caption{
Cumulative Costs
}
\label{fig:c123-cumul}
\end{figure}

We also plot cumulative cost for C1--C3 in Figure~\ref{fig:c123-cumul}; all models exhibit broadly similar cost growth over time.

\pagebreak 

\label{section--References}
\printbibliography

\clearpage
\section*{Appendix}

\subsection*{A. Decoupling of the Alternating Minimization Objective}
\label{app:altmin-decoupling}

We start from the objective
\[
f(U,V)=\|\mathcal{P}_\Omega(UV^\top - Y_{\mathrm{obs}})\|_F^2
      =\sum_{(i,j)\in\Omega} (u_i^\top v_j - Y_{ij})^2,
\]
where $u_i^\top$ is the $i$-th row of $U$ and $v_j^\top$ is the $j$-th row of $V$. If we fix $U$, we may reorganize terms by column:
\[
f(U,V)
= \sum_{j=1}^m \sum_{i:(i,j)\in\Omega}
   (u_i^\top v_j - Y_{ij})^2
= \sum_{j=1}^m \underbrace{\sum_{i\in I_j}
   (u_i^\top v_j - Y_{ij})^2}_{\text{depends only on } v_j},
\]
where $I_j=\{\,i:(i,j)\in\Omega\,\}$ is the set of observed rows in column $j$.
Thus each $v_j$ solves a separate least-squares problem:
\[
v_j
=\arg\min_{v\in\mathbb{R}^r}
   \sum_{i\in I_j} (u_i^\top v - Y_{ij})^2.
\]
Symmetrically, if $V$ is fixed, we reorganize by rows where letting $J_i=\{\,j:(i,j)\in\Omega\,\}$  each $u_i$ solves
\[
u_i
=\arg\min_{u\in\mathbb{R}^r}
   \sum_{j\in J_i} (u^\top v_j - Y_{ij})^2.
\]
Hence by fixing one factor we break the objective into small, independent least-squares
problems, one for each row or column. Alternating minimization repeatedly
solves these problems, updating $U$ and $V$ in turn.

\end{document}